\documentclass[twocolumn,preprintnumbers,amsmath,amssymb]{revtex4}

\usepackage{graphicx,graphics,float}% Include figure files
\usepackage{dcolumn}% Align table columns on decimal point
\usepackage{bm}% bold math

\begin{document}

\title{End-to-end entanglement in Bose-Hubbard chains}

\author{Jose Reslen}
\author{Sougato Bose}

\affiliation{Dept. of Physics and Astronomy, University College London, Gower Street, WC1E 6BT London, United Kingdom}%

\date{\today}

\begin{abstract}

We study the ground state as well as the dynamics of chains of
bosons with local repulsive interactions and nearest-neighbour
exchange using numerical techniques based on density matrix
decimation. We explore the development of entanglement between the
terminal sites of such chains as mechanisms are invoked to
concentrate population in these sites. We find that long-range
entanglement in the ground state emerges as a result of transfer 
taking place across the length of the whole chain in systems with appropriate
hopping coefficients. Additionally, we find appropriate perturbations 
to increase the entanglement between the end sites above their ground
state values.

\end{abstract}

\maketitle

\section{INTRODUCTION}

\begin{figure}
\includegraphics[width=0.48\textwidth, angle=-0]{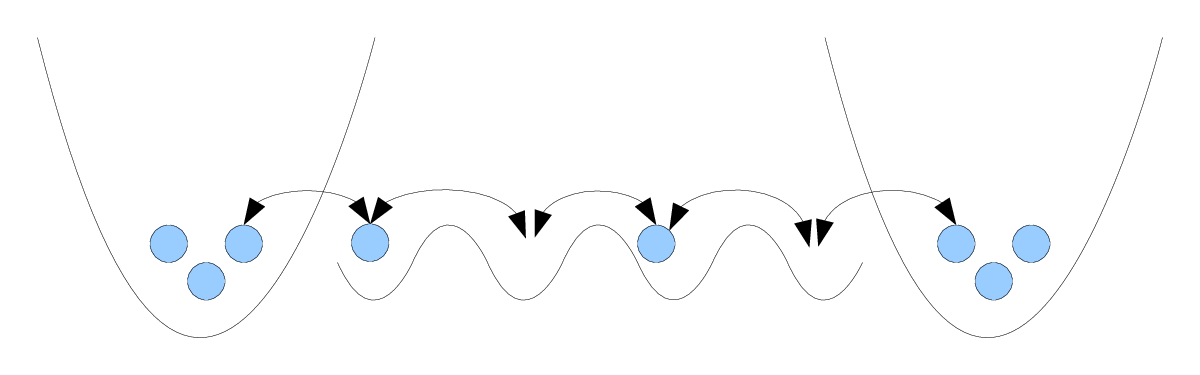}
% Here is how to import EPS art
\caption{\label{fig:zero}  Potential realization of a Bose-Hubbard
chain in an optical lattice. The above finite lattice can be
realized in one cell of a superlattice. The end sites of the lattice
are depicted differently in the figure as we will require them to
have vanishing repulsion in our study.}
\end{figure}

As constant experimental developments expand our possibilities of
materializing a new generation of computing devices, the need for a
better understanding of quantum phenomena becomes more important.
Outstanding emphasis has been placed on the concept of entanglement
as, for example, a resource for quantum information processing and
in general as a tool to study the contrast between classical and
quantum physics. One fundamental problem that is now the subject of
intensive research is the emergence of quantum correlations between
distant sites of a quantum system. For most many-body systems this
is notoriously difficult and entanglement between distant sites can
only be achieved through clever engineering \cite{lorenzo} or 
non-equilibrium dynamics \cite{Eisert,Yung,Tsomokos,Oriol,Spiller}, where, in
the latter case, either intricate methodology is required or the amount of
entanglement between the ends, though finite, is not substantial in amount. 
Typically, in quantum many-body systems, even when total correlations
are long range, the entanglement between individual
constituents such as spins is extremely short range, such 
as between nearest or next to nearest neighbours \cite{amico}. Thus it is
already interesting when in some many-body system an entanglement between
its farthest components, such as the spins or harmonic oscillators at the very
ends of an open chain, can be generated and even more interesting when it is
substantial in amount.
Most of the work done in this field has been carried out in light-matter systems and
spin chains, but more recently, the development on new numerical
techniques and algorithms \cite{vidal1} has opened the possibility
of addressing more challenging scenarios such as chains of interacting
bosons \cite{Oriol,Moeckel,Daley,several}. Usually, where bosonic chains
have been studied, though true quantum correlations have indeed been
found to emerge between distant individual sites, they are
generically not ``substantial'' in magnitude \cite{Eisert,Plenio}. 
Thus substantial entanglement between distant sites of a Bose-Hubbard
model, even if with some reasonable engineering of the Hamiltonian,
should be of great interest, not only because of the uncommon nature of such
long distance entanglement, as motivated above, but also because from the
practical point of view this entanglement is known
to be ``distillable", i.e., convertible from an impure (mixed state) to a pure
useful form through local actions. Density matrices whose
partial transposes display negative eigenvalues are distillable \cite{Stockton}
-- so if one finds a mixed entangled state with the above property then, one can
already claim, at least {\em in principle}, to have a resource for quantum 
communication. 

Chains of bosons are more often than not well described by the
Bose-Hubbard model \cite{Fisher,Cirac,Roth}. In this model bosons can hop
between neighbouring sites while undergoing local repulsion when
several bosons occupy the same site. The development in
atom cooling techniques has led to a significant rise in the
amount of experiments related to the Bose-Hubbard model. The 
transition from a superfluid to a Mott insulator has been verified in 
numerous experiments of cold atoms in optical lattices as for example in  
Refs. \cite{Bloch,Cirac,Chang,Stoferle,Muller}, just to mention a few.
As a result, physical realizations of quantum systems displaying
key features such as long range entanglement is becoming more and
more feasible. Motivated by these advances, in this paper we analyse from a
fundamental point of view the emergence of long-range quantum
correlations in Bose-Hubbard chains (Fig. \ref{fig:zero}). There has
been one previous investigation on long range entanglement in a
Bose-Hubbard chain where the dynamics is effectively reducible to a
single particle propagation in a lattice \cite{Oriol}. Our study is,
however, irreducibly a many-particle situation and can even result
in more entanglement than what a single particle hopping in a
lattice can ever do. It is also pertinent to point out that there
has been much interesting work on the various forms of entanglement
contained in Bose gases \cite{Simon,Anders}, but here we concentrate
on a specific type of entanglement, namely that between two
individual sites at the very ends. If implemented in the laboratory, the 
entanglement obtained from the scheme presented in this work, as it is
distillable, would provide a resource for quantum information applications. 
In addition to the end-to-end entanglement, which has been
motivated above, in this paper we will also present some results for 
the entanglement between a couple of other bi-partite partitions of 
the chain to fully appreciate the distribution of entanglement 
between parts of a finite  chain as the strength of the local 
repulsion between bosons is varied.

\section{THE HUBBARD MODEL: CHARACTERISTICS AND METHODS}

We assume a chain of size $N$ with $M$ bosons, nearest-neighbour
hopping and open ended boundary conditions governed by a
Bose-Hubbard Hamiltonian with variable coefficients,

\begin{equation}
\small{
\hat{H} = \sum_{k=1}^N {\frac{U_k}{2} \hat{a}^{\dagger}_k \hat{a}_k (\hat{a}^{\dagger}_k \hat{a}_k -1)} - \sum_{k=1}^{N-1} {J_k (\hat{a}^{\dagger}_{k+1} \hat{a}_k  + \hat{a}^{\dagger}_k \hat{a}_{k+1}  )}
}
\label{eq:one}.
\end{equation}

Constants $U_k$ and $J_k$ account for the on-site repulsion and
hopping respectively while the annihilation and creation operators
$\hat{a}^{\dagger}_k$ and $\hat{a}_k$ obey the usual commuting rules
$[\hat{a}_l,\hat{a}^{\dagger}_k]=\delta_k^l$
 and $[\hat{a}^{\dagger}_k,\hat{a}^{\dagger}_l]=[\hat{a}_k,\hat{a}_l]=0$.
Physical realizations of
this Hamiltonian include experiments
where remarkable control of the ratio $U/J$ can be achieved \cite{Bloch,Muller,Cirac}.
In a typical experiment, atoms are cooled in an optical lattice of retroreflected diode 
lasers and then transferred into a magnetic trap where further cooling is to take
place. This creates an arrangement of atoms where the resulting optical potential
depths $V_{x,y,z}$ are proportional to the laser intensities and can be expressed
in terms of the recoil energy $E_{R}=\frac{\hbar k^2}{2 m}$ with $m$ the atomic mass and
$k$ the wavelength number. To prepare 1D arrays two lattice lasers are given
high intensities in such a way that hopping can only efficiently take place across 
one axis \cite{Stoferle}. In terms of experimental parameters, hopping and repulsion
coefficients are given by $J= A \left( \frac{V_{0}}{E_{R}} \right )^B e^{-C\sqrt{V_{0}/E_{R}}} E_{R} $	 
and $U=\frac{2 a_{s} E_{R}}{d} \sqrt{\frac{2 \pi V_{\perp}}{E_{R}}} \left( \frac{V_0}{E_{R}}  \right)^{\frac{1}{4}} $
where $V_0$ is the axial lattice depth, $V_{\perp}$ the depth of the lattice
in the transverse directions, $a_s$ the $s$-wave scattering length, $d$ the lattice
spacing and $A$, $B$ and $C$ fixed constants \cite{Rey,Danshita}. Spatial variations in $U$ and $J$ 
can also be implemented using detuned lasers sent through specific sections of 
the lattice as reported in Ref. \cite{Chang}. 

From the Heisenberg equations of motion we know $\frac{d \hat{\alpha}_k}{d t} =
i e^{-i t \hat{H}} [\hat{a}^{\dagger}_k,\hat{H} ]  e^{i t \hat{H}}$,
where $\hat{\alpha}_k = e^{-i t \hat{H}} \hat{a}^{\dagger}_k  e^{i t
\hat{H}}$. In the general case non-linearities induced mostly by repulsive
terms keep us from getting
reliable expressions for the $\hat{\alpha}_k$ in terms of time. One
fortunate instance is the repulsionless case where all $V_k$ go to
zero and we are left with

\begin{equation}
\small{
\frac{d}{d t}
\left(
\begin{array}{c}
\hat{\alpha}_1 \\
\hat{\alpha}_2 \\
\hat{\alpha}_3 \\
\vdots \\
\hat{\alpha}_N \\
\end{array}
\right)
=
-i\left(
\begin{array}{ccccc}
0 & J_1 & 0 & \cdots & 0 \\
J_1 & 0 & J_2 & \cdots & 0 \\
0 & J_2 & 0 & \cdots & 0 \\
\vdots & \vdots & \vdots & \ddots & J_{N-1} \\
0 & 0 & 0 & J_{N-1} & 0 \\
\end{array}
\right)
\left(
\begin{array}{c}
\hat{\alpha}_1 \\
\hat{\alpha}_2 \\
\hat{\alpha}_3 \\
\vdots \\
\hat{\alpha}_N \\
\end{array}
\right)}
\label{eq:three}.
\end{equation}

One can of course assume that all the coupling constants are equal, 
which corresponds to the usually studied, and perhaps the most natural, setting  of the 
Bose-Hubbard model. To contrast with this, one can also study a setting
with non-uniform couplings to investigate whether better end to end entanglement
can be obtained by appropriately engineering the couplings. As an example we will choose a  non-uniform hopping distribution that matches an angular
momentum representation of length $j=\frac{N-1}{2}$, namely $J_k =
\frac{\lambda}{2}\sqrt{k(N-k)}$ ($\lambda =2$ for all the simulations
presented in this work). The above couplings, when present
in spin chains, are known to facilitate a perfect quantum state
transfer \cite{PT}, which has been developed in context of the idea
of using spin chains to convey quantum states \cite{Bose1}. 
In a Hubbard model, the above coupling profile will transfer particles
perfectly from one end of the chain to the other \cite{Plenio} (however, 
the aim of this paper is different, namely generating entanglement 
between the ends). Moreover, because we are interested in studying 
end-to-end entanglement (EEE), hopping distributions are known to 
play a crucial role as it is particle tunnelling what determines 
how distant places get correlated with each other. So, in this work 
we compare results from two different hopping profiles, namely, 
the well known constant hopping (CH) $J_{k}=1$ 
and the above introduced perfect transmission hopping (PTH). 
The purpose in bringing in PTH is to explore what kind of physics 
is shown by a hopping profile with perfect transmission properties 
in situations where transport is either absent or not directly involved. 
The natural question in this context is whether perfect transmission 
properties, which involve dynamical synchronization at long scales, 
are in any way linked to the onset of quantum correlations among
distant sites of a boson chain.

On the other hand, when repulsive terms in Eq. (\ref{eq:one}) are considered
we apply numerical methods.
Here we make use of density-matrix techniques as presented in
\cite{vidal1}, also known as {\it TEBD}, combined with conservation
and symmetry properties associated with the Hamiltonian.
Briefly, {\it TEBD} consists in writing the state in terms of
canonical coefficients that are closely related to Schmidt vectors.
These coefficients are updated through successive applications of
semi-local unitary operations that correspondingly only modify
coefficients associated with the sub-spaces on which they act.
This fact is then skilfully exploited to work out an
efficient way of simulating both real and complex evolution. 
In addition to the standard set of coefficients $\Gamma$ and
$\lambda$ employed in the canonical representation of the state, 
we also retain the number of particles associated with every 
Schmidt vector in the decomposition. Memory storage is handled 
by making use of allocatable types and pointers in FORTRAN 95, which suits
the mechanics of TEBD where the dimension of the Hilbert space 
is updated adaptively. Usually, the TEBD coefficients
that determine the local properties of the chain are presented
with explicit reference to every local dimension, i.e. 
$\Gamma_{\alpha \beta}^{[n]i}$, where $i=1,..,M$, being $M$ the
number of total number of bosons in the chain. However, the only
relevant component $i$ above is completely determined by the number
of particles $m_{\alpha}$  and $m_{\beta}$ associated with Schmidth vectors 
$|\alpha\rangle$ and $|\beta\rangle$ in the expression 
$|\psi \rangle = \sum_{\alpha,\beta,i} \Gamma_{\alpha \beta}^{[n]i} |\alpha \rangle |i\rangle |\beta \rangle$ through the relation $m_{\alpha}+m_{\beta}+i=M$.
Consequently, truncating the local Hilbert space as a way of attenuating
the memory requirement of the simulation is not necessary. Similarly, 
we use dynamical
allocation and deallocation to store the canonical representation,
which allows an efficient handling of computer memory. The maximum
number of Schmidt coefficients $\chi$ as well as the size of other 
elements relevant to the simulation are updated at run time. 
After every updating step we retain all the Schmidt coefficients 
greater than $\lambda_{1}^{i} \times 10^{-14}$, where $\lambda_{1}^{i}$ is the 
greatest coefficient at site $i$. In ground state simulations, $\chi$ saturates and 
remains quite fixed until the state converges (Insets in Fig. \ref{fig:n3}). 
For real time simulations, on the other hand, this convenient 
saturation does not manifest as strongly as in the ground state case 
and in some cases, specially for long simulations, truncation in the 
number of coefficients is important as a mean of sustaining the
simulation efficiency. In our program we use 
second order Trotter expansion with time steps $\delta t =\frac{10^{-3}}{N_e}$,
where $N_{e}$ is the effective length of the chain, which is to be
taken just as the number of sites in the chain unless something else
be specified. Here we limit ourselves to Hamiltonians with symmetric distributions
of parameters with respect to the centre of the chain and with an
even number of sites and bosons. Additionally, all the simulations
presented in this work correspond to $M=N$. Therefore, the system can
be genuinely simulated by focusing on the canonical coefficients of
half of the chain plus the interaction between complementary half-chain
blocks. The algorithm is applied recursively to simulate 
$|\psi_G \rangle = lim_{\tau \rightarrow \infty}
 \frac{e^{-\tau \hat{H}}|\psi_0 \rangle}{||  e^{-\tau \hat{H}}|\psi_0 \rangle ||}$,
until convergence to the ground state $|\psi_G \rangle$ is accomplished
according to the criterion
$|1-\langle \psi(\tau) | \psi(\tau+\delta \tau) \rangle |<10^{-14}$.
Ground state simulations in repulsionless chains agree well with 
the exact ground state of Hamiltonian (\ref{eq:one}), which for
PTH chains reads

\begin{equation}
\small{
\left ( \sum_{m=-j}^j {\frac{(-1)^{j+m}}{2^j}\sqrt{\frac{(2j)!}{(j-m)!(j+m)! }} \hat{a}_{m+j+1}^{\dagger}  }    \right )^M \prod_{l=1}^N |0_l \rangle
\label{eq:five},
}
\end{equation}

with energy $E_G = -\lambda j N$. Similar expressions for the ground
state of CH chains can be found elsewhere. Additionally, real-time 
simulations coincide with available theoretical results \cite{Plenio,Hartmann,Oriol}.

As it was stated from the introduction of TEBD and 
further verified in several numerical studies \cite{TEBD-study}, efficient simulations 
can be carried out as long as the state can be represented using 
reasonable low $\chi$, i.e., correlations are not too strong. 
As a reference one could take the infinity Hubbard model with unit filling, 
where such regime would correspond primarily to the Mott insulator 
phase where according to mean field theory $\frac{U}{z J } > 5.8$ 
with $z=2 d$ the number of nearest neighbours \cite{Fisher,Cirac}. 
Another efficient simulation scenario corresponds to the 
case when, although hopping constants are strong and superfluid features
dominate, the Hamiltonian structure remains close to integrability, e.g.
$\frac{U}{J}\ll 1$, since usually integrable dynamics takes place 
in subspaces. When the case is neither of the above mentioned, 
as it is for most of the simulations presented from now on, one can still see 
that $\chi$ often remains  well below the maximum allowed in 
the Hilbert space.

\section{RESULTS}

\subsection{Ground state}

\begin{figure}
\includegraphics[width=0.34\textwidth, angle=-90]{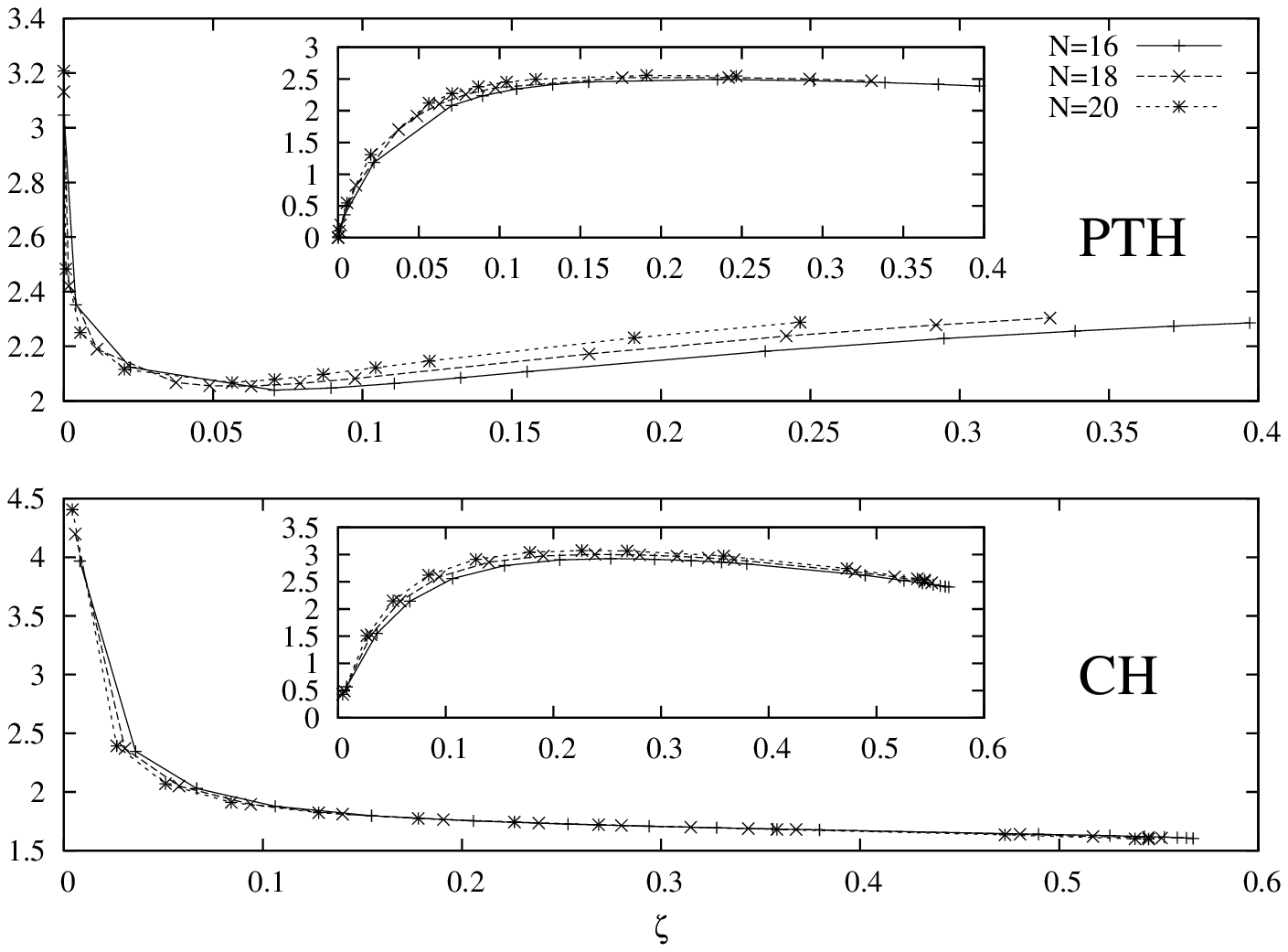}
% Here is how to import EPS art
\caption{\label{fig:n1}
Entanglement between both halves of the chain (main plots) and entanglement
between the ends and the rest of the system (Insets). In PTH chains, when
repulsion is sufficiently strong tunnelling starts to take place
throughout the whole chain, 
enhancing the amount of entanglement in the system. CH chains, on the other
hand, display an entanglement register that can be understood as being
originated from short scope hopping.
}
\end{figure}

\begin{figure}
\includegraphics[width=0.34\textwidth, angle=-90]{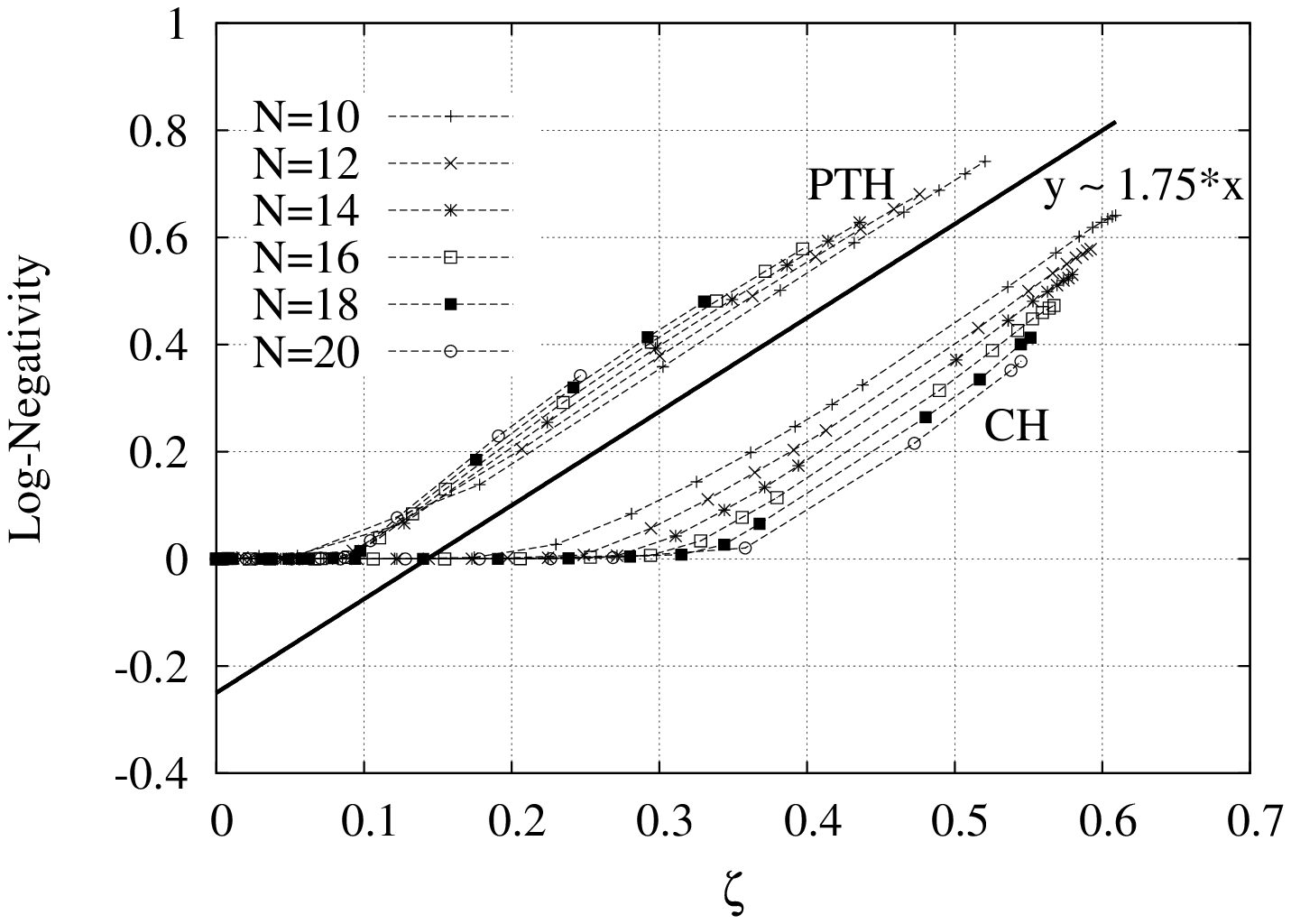}
% Here is how to import EPS art
\caption{\label{fig:one}  
The emergence of EEE
in the chain as the on-site repulsion or equivalently the fractional population at the
end sites $\zeta$ is increased. Linear behaviour can be observed independently of the chain size. 
Also, EEE in chains with PTH lingers on when
the size of the chain is increased, in contrast to EEE in CH chains.
}
\end{figure}

In order to calculate the entanglement between the ends of the
chain, we need to get the corresponding reduced density matrix.
This task is by itself challenging, as the quantum state is 
given in terms of canonical coefficients associated with TEBD and
not in the standard basis. Similarly, the computation of such density matrix 
is greatly improved in terms of speed and memory when exploiting 
number conservation. Once this density matrix is obtained, we have to 
quantify the entanglement using a measure which is appropriate 
for mixed states of two
arbitrary dimensional quantum systems. So we choose the {\em
logarithmic negativity} as defined in Ref. \cite{LogN}.
Moreover, to measure the entanglement between complementary
subsystems \cite{Latorre}, we use {\em von Neumman entropy} 
$S=-tr(\hat{\rho_A}log_2 (\hat{\rho_A}) )$ where $\hat{\rho}_A$
represents the reduced density matrix of subsystem $A$.
In this work we focus on chains with unit filling and open
boundary conditions. Here we show results against the 
fraction of particles present on the ends 
$\zeta= \frac{ 2 \langle \hat{a}^{\dagger}_1 \hat{a}_1  \rangle}{M}$
which allows a convenient depiction of the system phenomenology for
different chain sizes. $\zeta$ varies according to the intensity of 
the repulsion in intermediate sites. When repulsion is small, 
few particles remain on the ends and $\zeta \approx 0$. When 
repulsion is strong the amount of particles on the ends is maximum 
and $\zeta = \frac{1}{2} + \frac{1}{M}$.

Irrespective of the hopping profile the ground state
of repulsionless chains is highly superfluid,
although with most of the hopping taking place in the central part,
leaving the terminal positions nearly unoccupied.
In this repulsionless regime entanglement is strong, but confined around 
the centre. For example, in these circumstances, it is strong between the left
and right halves of the chain. When repulsion in intermediate positions is increased,
particles are forced to hop through longer distances and
correlations develop at longer scales. Similarly, 
bosons accumulate on the terminals according to the
strength of the interaction in intermediate places. Notice
that on account of the repulsionless ends, particles can 
always find channels to hop, no matter how intense the 
repulsion in intermediate sites. Small repulsion leads to 
a decrease of entanglement between both halves and an increase 
of entanglement between both ends and the rest of the system, 
an indication that entanglement is being spread along the chain 
following the particle distribution profile. When repulsion is
increased even more, the terminals start getting 
macroscopically occupied while the average number of bosons
in intermediate sites falls asymptotically towards $\frac{1}{2}$,
which means that strong tunnelling renders the quantum
state of intermediate places into a state that involves superpositions of
kets corresponding to one or zero bosons. In this case correlations among 
places near the ends and in opposite sides of the chain are strongly
enhanced, in contrast to the correlations in
the centre of the chain. This can be seen in the inset plots
of Fig. \ref{fig:n1} where von Neumman entropy
between both ends and the rest of the system
slightly comes down after the original redistribution
of entanglement mentioned above occurs. The emergence of EEE
in the chain is shown in Fig. \ref{fig:one}, which shows that 
as repulsion in the centre becomes intense, the terminals will tend
to entangle with each other more strongly than with intermediate 
sites. 

Once the conditions for the emergence of EEE are given, what determines 
how much entanglement can be generated is the hopping {\it scope}. 
Indeed, what is being really adjusted when repulsion is turned on is the
effective travelling length of bosons along the chain.
As repulsion in the middle is tuned, bosons get to hop through longer 
distances and thereby their become more and more delocalized. 
The behaviour displayed
by von Neumman entropy between both halves in consistent with
a regime in which entanglement is being continuously redistributed
across the system as a result of increasing hopping scope, 
but the fact that there is no turning point in Fig. \ref{fig:n1}
for CH indicates that such hopping never takes place across the complete 
length of the chain. Nevertheless, such hopping is enough to 
induce EEE at finite $N$, but not in the thermodynamic limit
since EEE dies down against increasing chain size. In PTH chains,
on the other hand, there exist a point in which hopping scope actually 
embraces the whole chain, which facilitates the particle transport
from one end to the other, making the proportion of
particles being held on the terminals useful for entanglement.
Once this long scale hopping takes over, increasing the repulsion 
in intermediate sites reinforce end-to-end exchange 
and induces an increase in the amount of entanglement 
contained in the system, as can be seen in Figs. \ref{fig:n1} 
and \ref{fig:one}. At the same time, EEE begins to show a linear 
dependence with $\zeta$ with a slope that is independent of the 
system size. In this case EEE remains even when the size of the
chain is augmented, which means that long range correlations
will manifest themselves in the thermodynamic limit.

\subsection{Dynamics}

\begin{figure}
\includegraphics[width=0.34\textwidth, angle=-90]{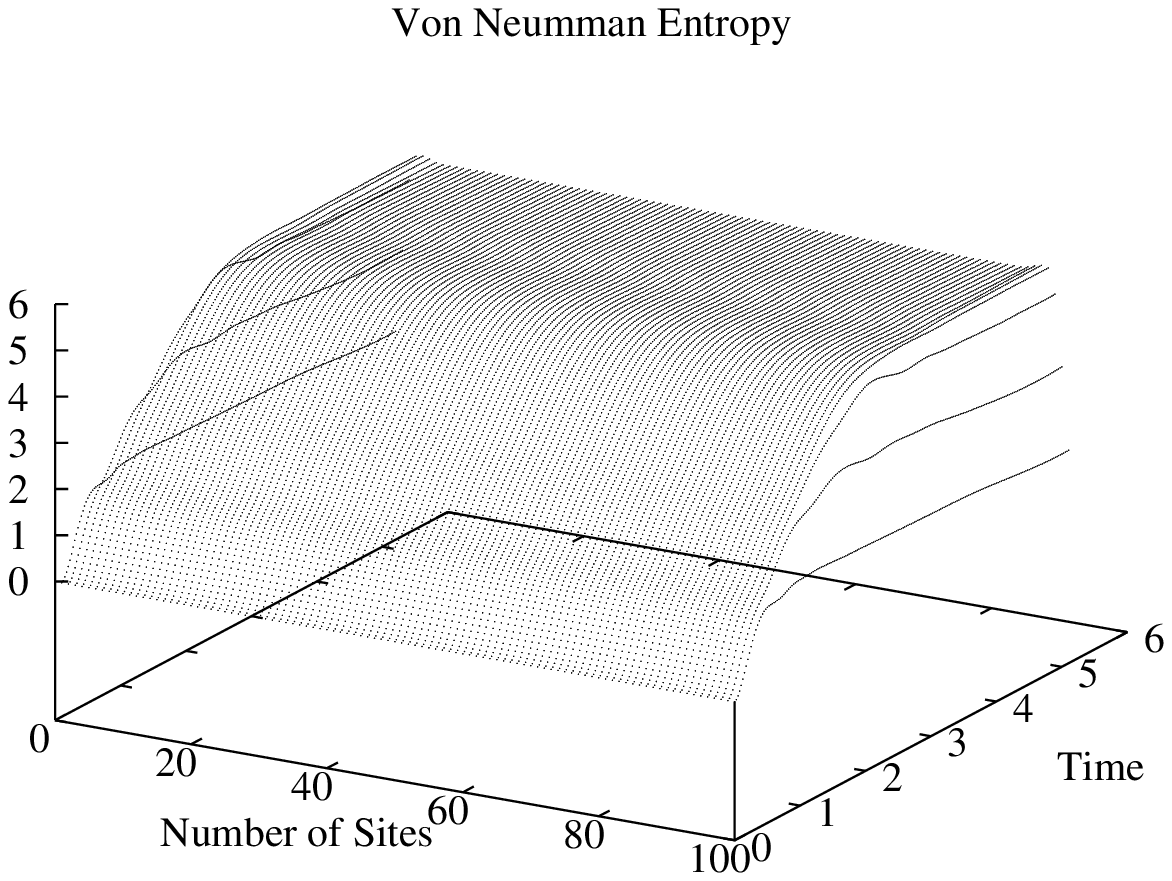}
% Here is how to import EPS art
\caption{\label{fig:21}
Entanglement between complementary blocks in a chain with $N=100$
and $U/J=0.4$. In this simulation $\chi=50$ and 
$\delta t = 5 \times 10^{-5}$. Entanglement saturates
uniformly except close to the ends, where it displays smaller
saturation values.
}
\end{figure}

\begin{figure}
\includegraphics[width=0.34\textwidth, angle=-90]{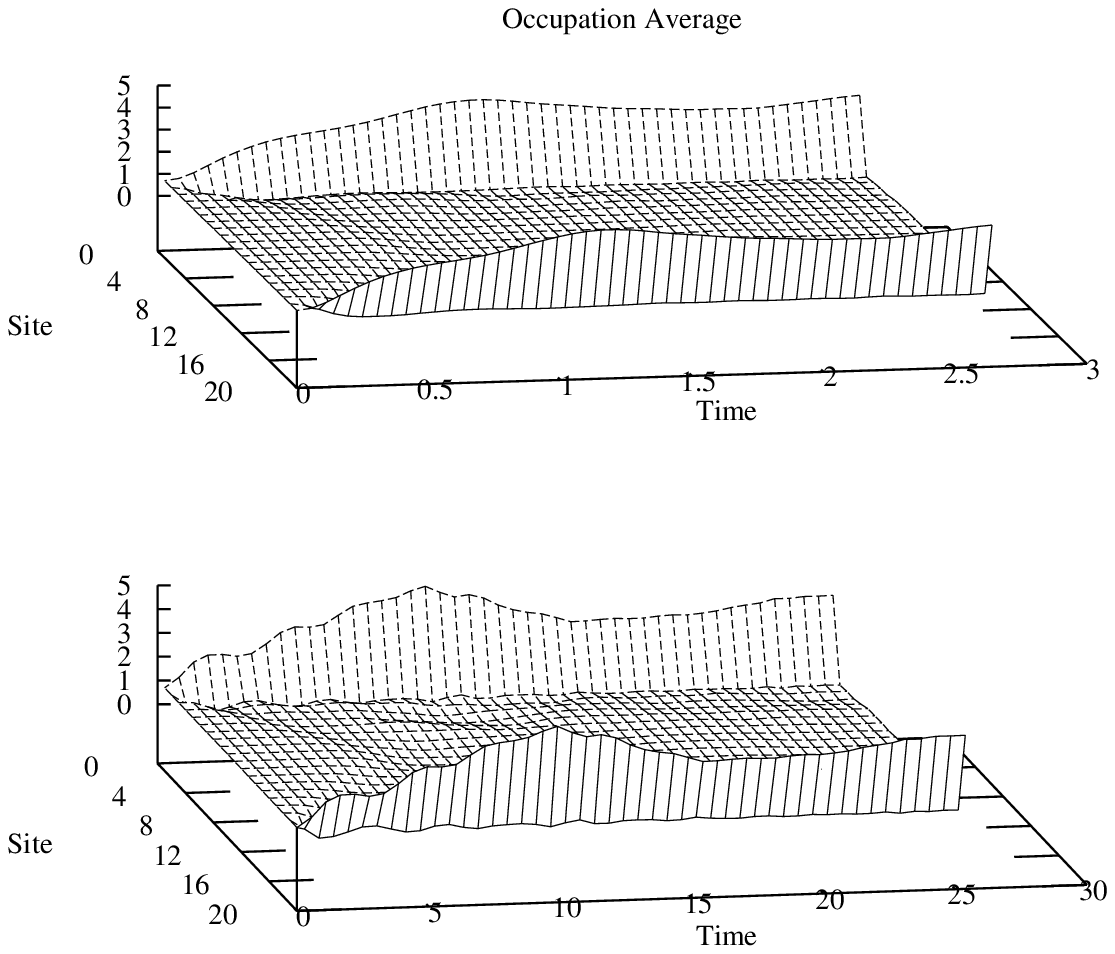}
% Here is how to import EPS art
\caption{\label{fig:6}
Average number of bosons in chains with $N=20$ and $M=20$. 
$\chi=50$. $U=100$ for PTH (upper plot) and $U=200$ for CH (lower plot). 
Stronger repulsion constants are necessary in PTH chains in order
to maintain intermediate sites fermionized so that no more than
one particle can be at these sites. The initial
state is a Mott insulator with one boson per site. At
the beginning bosons start to migrate towards the ends,
leaving the intermediate of the chain with half a boson 
per site on average. As a result of the hopping, EEE
emerges after some time (Fig \ref{fig:5}). For longer 
times slow oscillations can be seen, but the ends remain
well populated.  
}
\end{figure}

\begin{figure}
\includegraphics[width=0.34\textwidth, angle=-90]{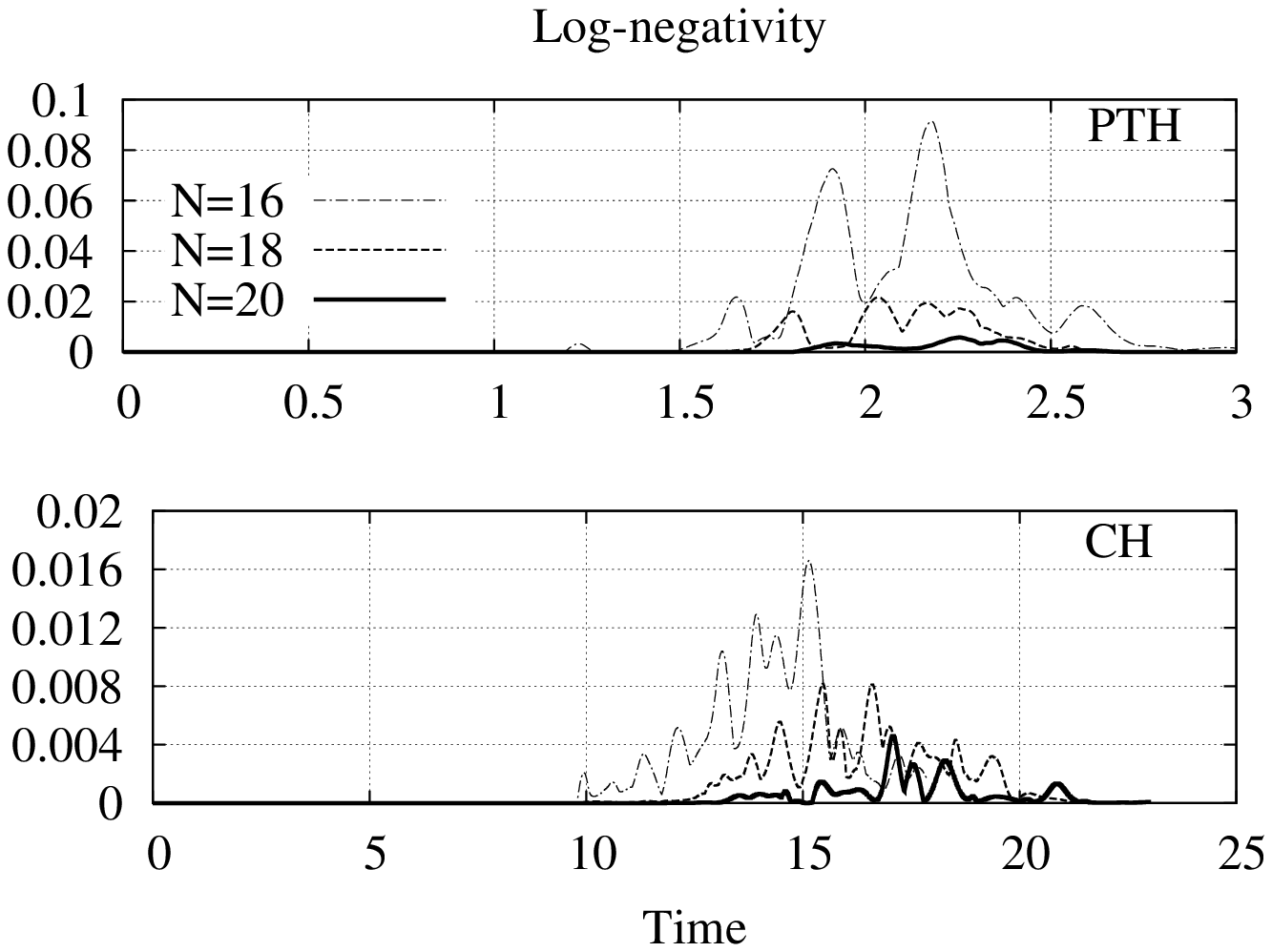}
% Here is how to import EPS art
\caption{\label{fig:5}
EEE as a function of time for different chain sizes. $\chi=50$. 
$U=100$ for CH and $U=200$ for PTH. The purpose
is to restrain particle accumulation anywhere but the terminal sites. 
Entanglement in PTH chains is not only bigger
but it also arises at early times than in CH chains. In both
plots Log-negativity decreases with the size of the chain.
Also, entanglement emerges well after bosons migrate
to the ends (Fig \ref{fig:6}).
}
\end{figure}

Here we investigate how EEE entanglement shows up as a result
of dynamics in a chain initially prepared in a Mott insulating state. 
The typical behaviour of entanglement in chains with 
homogeneous repulsion and hopping constants is shown in Fig. \ref{fig:21}.
As hopping is bigger than repulsion, particle tunnelling 
generates rich dynamics and entanglement behaves non-trivially 
in contrast to the high repulsion regime. Entanglement saturation
is identical for all sites except those very close to the terminals, where 
saturation takes place at smaller values. Such boundary effects are
typical of open chains \cite{several,Kollath}. 
As a consequence, the ends are poorly entangled 
with the rest of system and between them as well. On the other hand,
dynamics in repulsionless chains is known to lead to a progressive
thermalization of reduced density matrices of each site
\cite{Cramer} and therefore any sort of two site entanglement is weak. 
In this sense, dynamic 
EEE in repulsionless chains is characteristically similar to that found 
previously in the ground state. As an alternative approach, we try the
same methodology already applied and set all intermediate
repulsion constants to high values in order to artificially create
a particle distribution favourable to EEE. The effect on the average
number of particles can be seen in Fig. \ref{fig:6}. The ends get
macroscopically occupied and tunnelling in intermediate places intensifies.
EEE arises some time after the ends have been occupied (Fig. \ref{fig:5}). 
In general, the bigger the chain the longer it takes for EEE to emerge.
For PTH, a natural time scale is determined by the
transmission period, namely $T=\pi$ given our particular choice of $\lambda$.
Entanglement arises after half of such period, just after information has got
to travel from one end of the chain to the other, but there is not enough
evidence that this is always the case, particularly for very long chains. 
Additionally, it is possible that entanglement will build up further for longer times, independently
of the hopping profile, but this effect is difficult to observe as 
long time simulations require additional computational resources \cite{Osborne,TEBD-study}. 

\subsection{Perturbation approach and detection}

\begin{figure}
\includegraphics[width=0.34\textwidth, angle=-90]{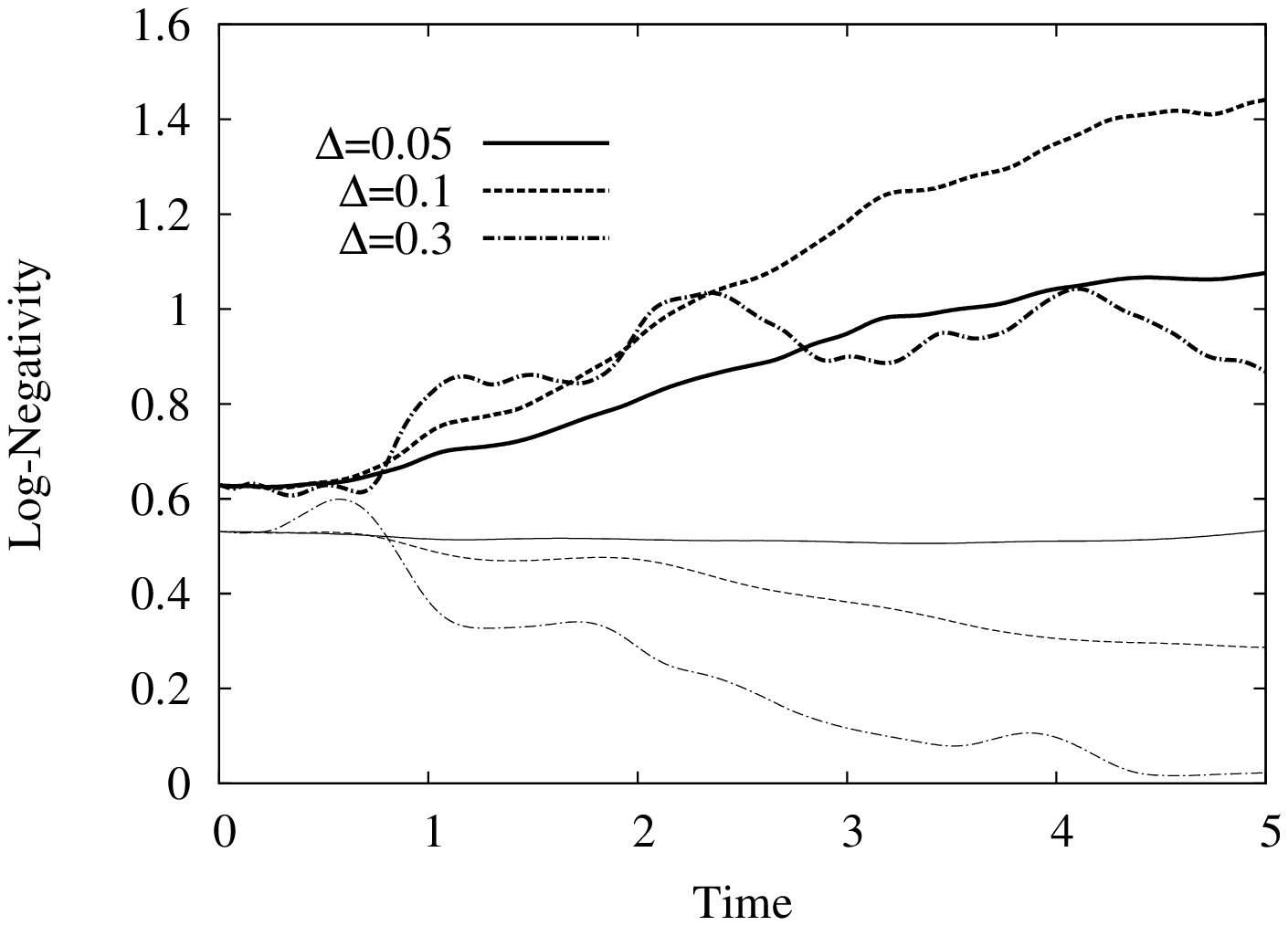}
% Here is how to import EPS art
\caption{\label{fig:n2}
EEE in a chain with $N=14$ and different values of $\Delta$ 
for both CH and PTH. PTH lines are thicker than their CH 
equivalents. EEE in the ground state can be improved as a 
result of dynamics in PTH chains. $\chi=200$ for PTH and 
$\chi=100$ for CH.
}
\end{figure}

\begin{figure}
\includegraphics[width=0.34\textwidth, angle=-90]{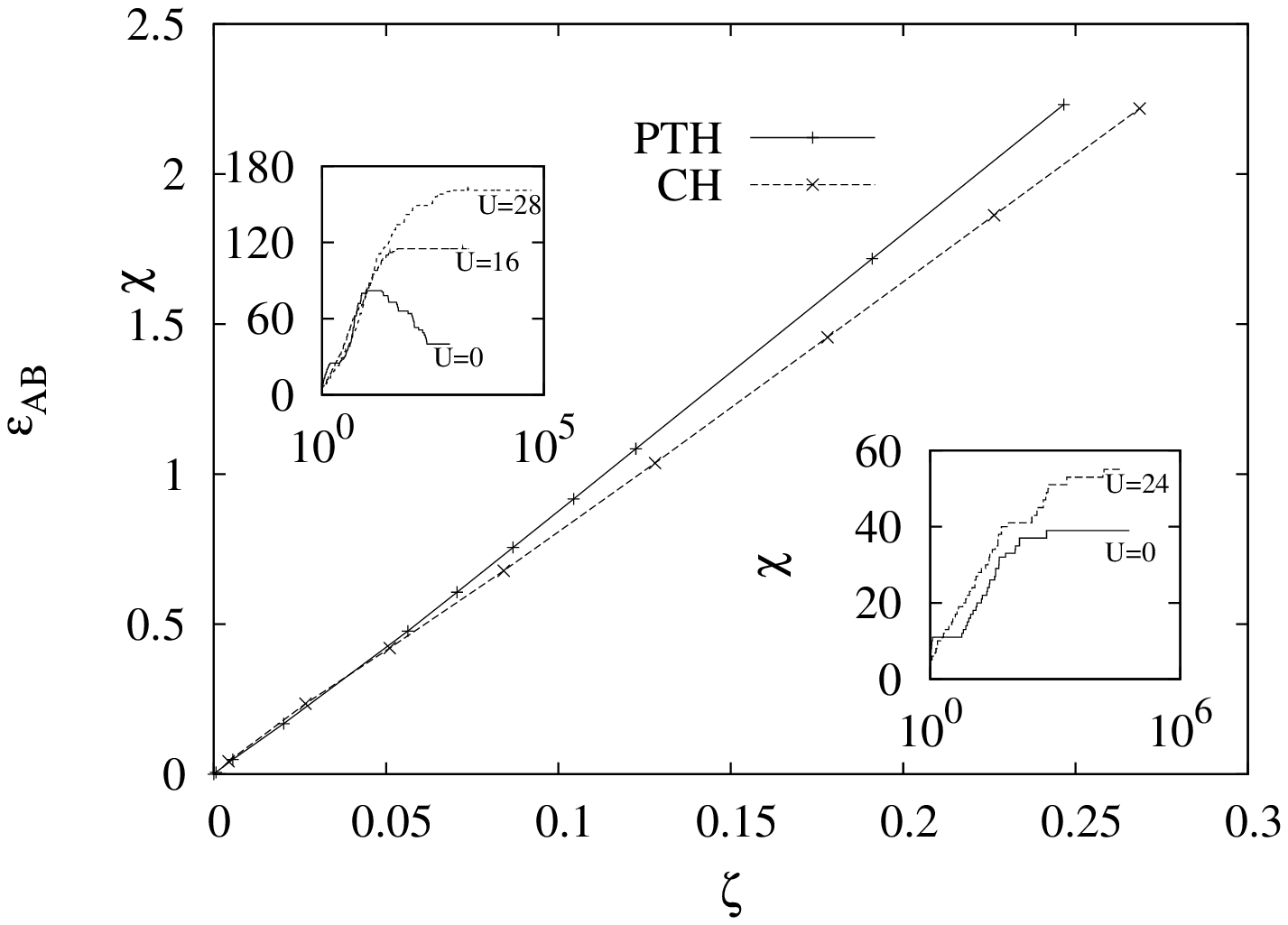}
% Here is how to import EPS art
\caption{\label{fig:n3}
Entanglement detection after bosons are 
sent through a 50:50 beamsplitter (see Ref. \cite{Vedral}). 
The original state is extracted from the ground state
of chains with N=20. Insets: Ground state convergence 
using the TEBD algorithm for PTH (left top) and CH (right bottom)
in chains with $N=20$.
}
\end{figure}

We would now ideally like to generate such an entanglement between
the ends where the logarithmic negativity exceeds unity (the maximum
value in a two qubit state) so that the possibilities of multiple
occupation numbers of the ending sites is utilized highly
beneficially. Looking at potential alternatives to the schemes
discussed before, we now assume that a chain initially prepared in
the ground state of a Hamiltonian with high intermediate repulsion
and highly fermionized intermediate positions 
is set to evolve under the action of a small perturbation potential
corresponding to a physical mechanism that internally pumps bosons
in and out the ends. The idea is to use the perfect transmission
properties at a perturbation level to generate entanglement as
a result of transport. The quantum state at any time can be written as

\begin{equation}
|\psi (t) \rangle =  e^{-i t (\hat{H} + \Delta \sum_{j=1}^N\hat{h}_j)}
|\psi_g \rangle,
\label{eq:twelve}
\end{equation}

where $|\psi_g \rangle$ is the ground state of $\hat{H}$, $\Delta$
is a small number and $\hat{h}_j$ represents a local operator acting
on site $j$. For consistency, we require $[\hat{H},\sum_{j=1}^N
\hat{h}_j]\ne 0$. Here we consider that the local perturbations can
be written as functions of the corresponding local number operator,
i.e. $h_j(\hat{N}_j)$, where $\hat{N}_j=\hat{a}_j^\dagger
\hat{a}_j$. We can establish a non-vanishing commutator between the
perturbation and the Hamiltonian by choosing, for every site but the
ends, $h_j(x) = k x$, where $k$ is the integer distance between site
$j$ and the closest end. Notice that this is equivalent to
considering a chain with spatially dependent chemical potential in
intermediate positions. Moreover, the primary factor determining
$h_{1,N}=h$ is the optimization of boson-transfer from intermediate
sites into the ends when an initial state $|\psi_g \rangle$ is set.
Supposing $\langle \hat{H} \rangle$ remains stable during
perturbation evolution and correlations between the Hamiltonian and
the perturbation can be ignored, we can use the number-of-boson
basis. In this way, performing average-energy balances among
sequential particle distributions underlying an optimal trajectory
we get

\begin{equation}
h(n_0 + \frac{1}{2}(\frac{N}{2}-k)+\frac{1}{2})-h(n_0 + \frac{1}{2}(\frac{N}{2}-k))=\frac{\Delta k}{2}
\label{eq:13},
\end{equation}

where $n_0$ is the average number of bosons allocated in one end at $t=0$.
After some algebraic manipulations we find $h(x) = c_2 x^2 + c_1 x$ with
$c_1 = (\frac{N-1}{2} + 2 n_0)\Delta$ and $c_2 = -\Delta$.
With this perturbation, the mean number of
bosons does not fluctuate much since the dynamics is still
governed by a Hamiltonian with strong repulsion coefficients in
intermediate sites and bosons remain squeezed into the ends. The 
leading mechanism in the evolution is the
equitable exchange of particles between the terminals and the
rest of the system. In addition, one can expect that such
exchange will improve the communication between distant places
and also the entanglement. Fig. \ref{fig:n2} shows log-negativity as
a function of time for different perturbation intensities. CH
results are shown for comparative purposes. In the initial stages of
evolution the dynamics is characterized by an increase of
quantum fluctuations on the ends of the chain accompanied by little
change in the average number of bosons. Hence, EEE
is enhanced while avoiding massive migration of bosons
towards the centre of the chain. Significantly, entanglement generation 
is improved not by adding interaction terms to the Hamiltonian but 
by adding a local perturbative potential.

Finally, we would like to comment about how this entanglement
can be verified experimentally. Once the atoms condense
in the ground state or after the dynamics has taken place,
the detection scheme presented in \cite{Vedral} could be used. Following
such a scheme, bosons on the ends are sent through a
50:50 beamsplitter and then the number of particles on the
outputs is counted. Entanglement can then be detected using
$\epsilon_{AB} = tr(\hat{a}^{\dagger}_c \hat{a}_c \hat{\rho} )  
- \frac{N}{2} = tr(\hat{a}^{\dagger}_1 \hat{a}_N \hat{\rho} )$ 
where $tr(\hat{a}^{\dagger}_c \hat{a}_c \hat{\rho} )$ is the 
number of particles in one output and $\hat{\rho}$ is the reduced 
density matrix describing the ends. $\epsilon_{AB}=0$ for
separable states and $\epsilon_{AB}>0$ for entangled states.
In this way, entanglement 
is characterized in terms of experimentally measurable parameters.
For illustrative purposes, we present in Fig. \ref{fig:n3} a 
plot of $\epsilon_{AB}$ obtained from the ground state of chains
with PTH and CH. We conclude that $\epsilon_{AB}$ correctly
determines whether the state is entangled. In order to illustrate 
how our simulations
converge to the ground state and the computational cost involved 
we also include plots of $\chi$ against the number of steps. Here we
emphasise that in our program $\chi$ defines the dimension of  
an allocatable array which is being continuously resized according
to the number of Schmidt coefficients in the canonical representation.

\section{SUMMARY}

We have studied the ground state entanglement as well as
the dynamical entanglement displayed by Bose-Hubbard chains for
several configurations of parameters. Our results indicate that
chains with strong repulsion coefficients in intermediate places are
suitable scenarios for the emergence of long scale hopping that 
can lead to a subsequent development in long-range quantum
correlations. Entanglement also emerges as a result of dynamics
in chains initially prepared in a Mott insulator state.
In addition, we showed that ground state entanglement
can be improved through perturbation dynamics. Irrespective of the
issue of entanglement, the behaviours displayed by the ground and
dynamical states that we have predicted for different repulsion
profiles should be interesting to verify in an experiment.
It would also be interesting to investigate how the transition 
from short scope to long scope hopping occurs. It could happen, and
our results provide some evidence of this, that this transition
is in fact a second order phase transition where EEE works as
the order parameter.

JR acknowledges an EPSRC-DHPA scholarship. SB acknowledges support
of the EPSRC, the EPSRC sponsored QIPIRC, the Royal Society and the
Wolfson Foundation. We thank T.S. Monteiro and C. Hadley for very
helpful comments.

\end{document}